\documentclass[twocolumn,preprintnumbers,amsmath,amssymb,superscriptaddress,footnote]{revtex4}
\usepackage{epsfig,bm}
\usepackage{ulem}
\usepackage{amsmath,color}

\begin{document}
\title{Core swelling in spherical nuclei: An indication of the saturation of nuclear density}
\author{W. Horiuchi}
\email{whoriuchi@nucl.sci.hokudai.ac.jp}
\affiliation{Department of Physics,
  Hokkaido University, Sapporo 060-0810, Japan}

\author{T. Inakura}
\affiliation{Laboratory for Advanced Nuclear Energy, Institute of Innovative Research, Tokyo Institute of Technology, Tokyo 152-8550, Japan}

\begin{abstract}
\begin{description}  
\item[Background] 
  Nuclear radius is one of the most important
  and  basic properties of atomic nuclei
  and its evolution is closely related
  to the saturation of the nuclear
  density in the internal region but the systematics of the nuclear radii for
  the neutron-rich unstable nuclei is not well known.
\item[Purpose]   Motivated by the recent interaction cross section measurement
  which indicates the $^{48}$Ca core swelling in the neutron-rich Ca isotopes,
  we explore the mechanism of the enhancement of the neutron and proton
  radii for spherical nuclei.
\item[Methods] Microscopic Hartree-Fock calculations
  with three sets of Skyrme-type effective interactions
  are performed for the neutron-rich Ca, Ni and Sn isotopes.
  The total reaction cross sections for the Ca isotopes
  are evaluated with the Glauber model
  to compare them with the recent cross section data.
\item[Results] We obtain good agreement
  with the measured cross sections and charge radii.
  The neutron and proton radii
  of the various ``core'' configurations are extracted from
  the full Hartree-Fock calculation and discuss the core swelling mechanism.
\item[Conclusions]
  The core swelling phenomena occur depending on the properties
  of the occupying valence single-neutron states
  to minimize the energy loss that comes from the saturation
  of the densities in the internal region,
  which appears to be prominent
  in light nuclei such as Ca isotopes.
\end{description}
\end{abstract}
\maketitle

{\it Introduction--}
Systematic studies of the nuclear density and radius give
us an insight not only into the nuclear structure
such as halo~\cite{Tanihata85},
neutron skin~\cite{Suzuki95},
and nuclear deformation~\cite{Takechi10}
but also into the nuclear matter properties~\cite{Chen10,Reinhard10,
  RocaMaza11,Kortelainen13,Inakura13,RocaMaza15}.
The electron scattering measurements, which precisely
determine the charge distribution of nuclei have unveiled
the nuclear saturation properties~\cite{deVries87}.
These studies have been extending
to the short-lived neutron-rich unstable nuclei, in which
the behavior of the nuclear density is not well known.
The size properties of the unstable nuclei have often been
extracted from the total reaction or interaction cross section
measurements since the discovery of the neutron halo structure
in $^{11}$Li~\cite{Tanihata85}, and have been reached
at the heavier two-neutron halo nucleus $^{22}$C~\cite{Tanaka10,Togano16},
and neutron-rich Ne~\cite{Takechi10} and Mg~\cite{Takechi14} isotopes.
Recently, a sudden increase of the interaction
cross section beyond $^{48}$Ca was reported
indicating the swelling of $^{48}$Ca core
in the Ca isotopic chain~\cite{Tanaka20}.
It has been recognized that the sudden enhancement
of the matter radius is due to nuclear deformation
~\cite{Takechi10,Minomo11,Minomo12,Sumi12,Horiuchi12,Watanabe14,Takechi14}
or weakly bound neutrons like
the halo structure~\cite{Tanihata85,Tanaka10,Takechi14}.
However, these Ca isotopes are known
to be spherical~\cite{Stoitsov03,Delaroche10,Tagami20},
and they are well bound in $^{42-51}$Ca, $S_n\sim 5$--6 MeV~\cite{Wang17}.
This may call a new mechanism for the enhancement of the nuclear size.

In this paper, we aim to uncover the core swelling mechanism in
light-medium to medium spherical nuclei and
discuss how the ``core'' nucleus
is developed as a function of the neutron number 
within a microscopic mean-field theory, that is,
Skyrme-Hartree-Fock (HF) theory.
We investigate even-even proton closed-shell isotopes, i.e.,
Ca, Ni, and Sn isotopes, which exhibit almost spherical shape,
and discuss the evolution of the size of the ``core'' nuclei.
The core swelling phenomena may appear in the isotopic chain
across the major shell or subshell.
For example, the abrupt interaction cross section enhancement
from $^{22}$O ($N=14$ subshell) to $^{23}$O   
was found in Refs.~\cite{Ozawa00, Kanungo11}.
The core swelling in O isotopes was theoretically
discussed using $^{16}$O plus few-neutron models in Ref.~\cite{Masui12}.
This study also helps formulate ``core'' plus few-nucleon models
towards describing the light-medium mass dripline
nuclei, e.g., $^{22}$C~\cite{Horiuchi06, Singh19},
and $^{29,31}$F~\cite{Singh20, Masui20}.

{\it Theoretical models--}
Here we perform the Skyrme-HF calculation
and give a brief description of its model setup.
The numerical code used in the present calculation is a revised
version of the code developed in Refs.~\cite{Inakura06, Horiuchi12}.
The ground-state wave function
is obtained by minimizing the following energy density
functional~\cite{Vautherin72}, $E[\rho] = E_N + E_C - E_\mathrm{cm}$,
using the imaginary-time method~\cite{Davies80}.
The nuclear energy $E_N$ is given by a
functional of the nucleon density $\rho_q(\bm{r})$,
the kinetic density $\tau_q(\bm{r})$,
and the spin-orbit-current density
$\bm{J}_q(\bm{r})$ ($q=n,\,p$).
The Coulomb energy $E_C$ among protons is
a sum of direct and exchange parts.
The exchange part is approximated by means of the
Slater approximation, $\propto \int \mathrm{d}\bm{r} \, \rho_q(\bm{r})^{4/3}$.
$E_{\rm cm}$ is the center of mass energy.
The single-particle wave function $\phi_i(\bm{r})$
is represented in the three-dimensional grid points.
All the single-particle wave
functions and potentials except for the Coulomb potential
are assumed to vanish outside the sphere of radius 20 fm.
For the calculation of the Coulomb potential, we follow
the prescription given in Ref.~\cite{Flocard78}.
In order to see the model dependence,
we employ three kinds of Skyrme parameter sets,
SkM$^\ast$~\cite{SkMs}, SLy4~\cite{SLy4} and SkI3~\cite{SkI3}.

In this study, for the sake of simplicity,
we only consider spherical configurations and
neglect the pairing correlation.
The pairing correlation makes it difficult to define
the ``core'' configuration because this induces
the fractional occupation probability.
It should be noted that for nonclosed subshell nuclei,
the HF calculation produces slightly deformed ground states, whereas
the HF+BCS and Hartree-Fock-Bogoliubov
calculations give spherical ground states
due to the pairing correlations~\cite{Stoitsov03,Delaroche10,Horiuchi16,Horiuchi17}.
To preserve the spherical symmetry,
we employ the filling approximation~\cite{Beiner75}.
Namely, when the Fermi level with angular momentum $j$
is occupied partially by $m$ particles
($m < 2j+1$), we consider
that $2j+1$ particles with the uniform occupation probability
$m/(2j+1)$ occupy the Fermi level.
We note that the effect of the pairing interaction
on the nuclear radii is small for a spherical ground state
as was shown in Ref.~\cite{Horiuchi16}.

{\it Results and discussions--}
First we compute the total reaction cross sections 
of Ca isotopes following the prescriptions given in Ref.~\cite{Horiuchi12}. 
The nucleon-target formalism in the Glauber theory~\cite{Glauber}
(NTG~\cite{NTG}) is employed for the evaluation
of the total reaction cross sections.
Inputs to this reaction theory are the neutron and proton densities
of projectile and target nuclei as well as the profile function
which describes the nucleon-nucleon scattering at forward angles.
The neutron and proton densities of the projectile nuclei are
obtained from the HF calculations.
For the target nucleus, $^{12}$C, we adopt
the harmonic-oscillator density~\cite{Ibrahim09}
whose width parameter is fixed so as to reproduce
the root-mean-square (rms) point-proton radius of $^{12}$C, 2.33 fm, extracted
from the charge radius data~\cite{Angeli13}.
Standard parameter sets of the profile function
are tabulated in Ref.~\cite{Ibrahim08}. Once they are set,
this model has no adjustable parameter and gives a nice description
in high-energy nucleus-nucleus collisions as demonstrated in many examples
~\cite{Horiuchi06, Horiuchi07, Ibrahim09, Horiuchi10, Horiuchi12, Horiuchi15}.

\begin{figure}[ht]
  \begin{center}
    \epsfig{file=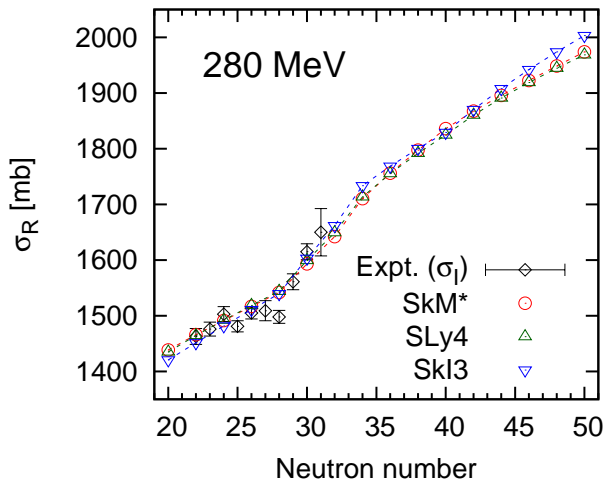, scale=1.5}
    \caption{Total reaction cross sections ($\sigma_R$) of Ca isotopes
      on a $^{12}$C target incident at 280 MeV/nucleon.
      The SkM$^\ast$, SLy4, and SkI3 interactions
      are employed for the HF calculations.
      Experimental interaction cross section ($\sigma_I$)
      data are taken from Ref.~\cite{Tanaka20}.}
    \label{rcs.fig}
  \end{center}
\end{figure}

Figure~\ref{rcs.fig} plots the calculated total reaction cross sections
of $^{40-70}$Ca isotopes on a $^{12}$C target at
the incident energy of 280 MeV/nucleon.
Recent experimental interaction
cross section data of $^{42-51}$Ca~\cite{Tanaka20}
are also plotted for comparison.
The calculated total reaction cross sections presented here
fairly well reproduce the recent experimental data~\cite{Tanaka20},
which confirm the validity of the calculated matter densities.
The interaction dependence of those adopted Skyrme parameter sets
on the cross sections are small
in the regions where the experimental data are available.
The SkI3 interaction predicts somewhat larger
cross sections for $N\gtrsim 30$ because SkI3 has larger
slope parameter of the equation of the state, the so-called $L$,
compared to others~\cite{Dutra12}, leading to more rapid growth of
the neutron skin thickness with respect to an increase of the
neutron number~\cite{Horiuchi17}.
The results of the neutron number $N>50$ are not shown
although the calculation still produces several bound ground states
with $N>50$ depending on the interaction employed.
Note that the heaviest Ca isotope found so far
is $^{60}$Ca~\cite{Tarasov18},
and the recent theoretical works predict the neutron dripline
is close to $^{60}$Ca~\cite{Hagen12},
and possibly at $^{70}$Ca~\cite{Newfcourt19},
and at $~^{64}$Ca~\cite{Tagami20}.  

\begin{figure}[ht]
  \begin{center}
    \epsfig{file=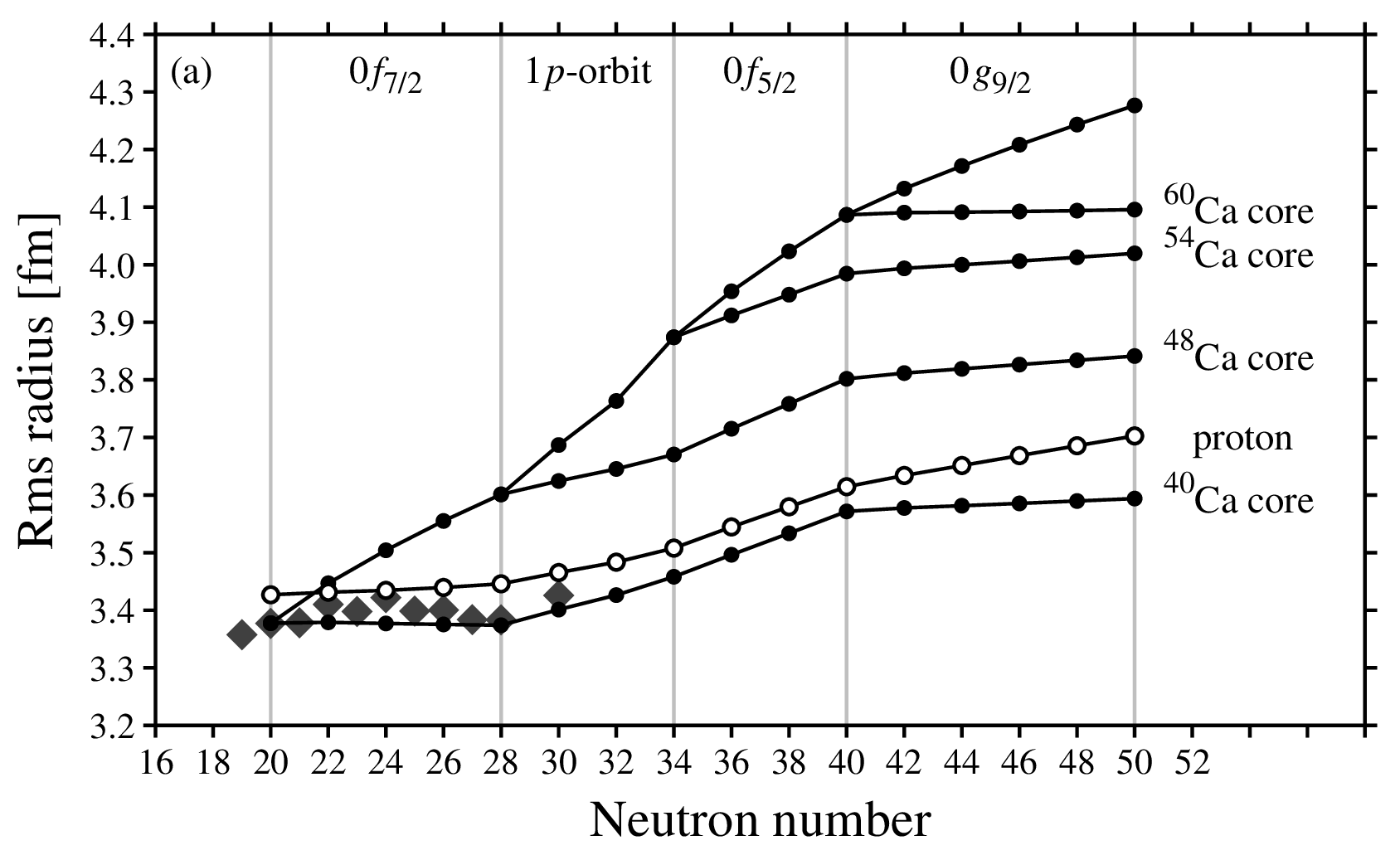, scale=0.3}
    \epsfig{file=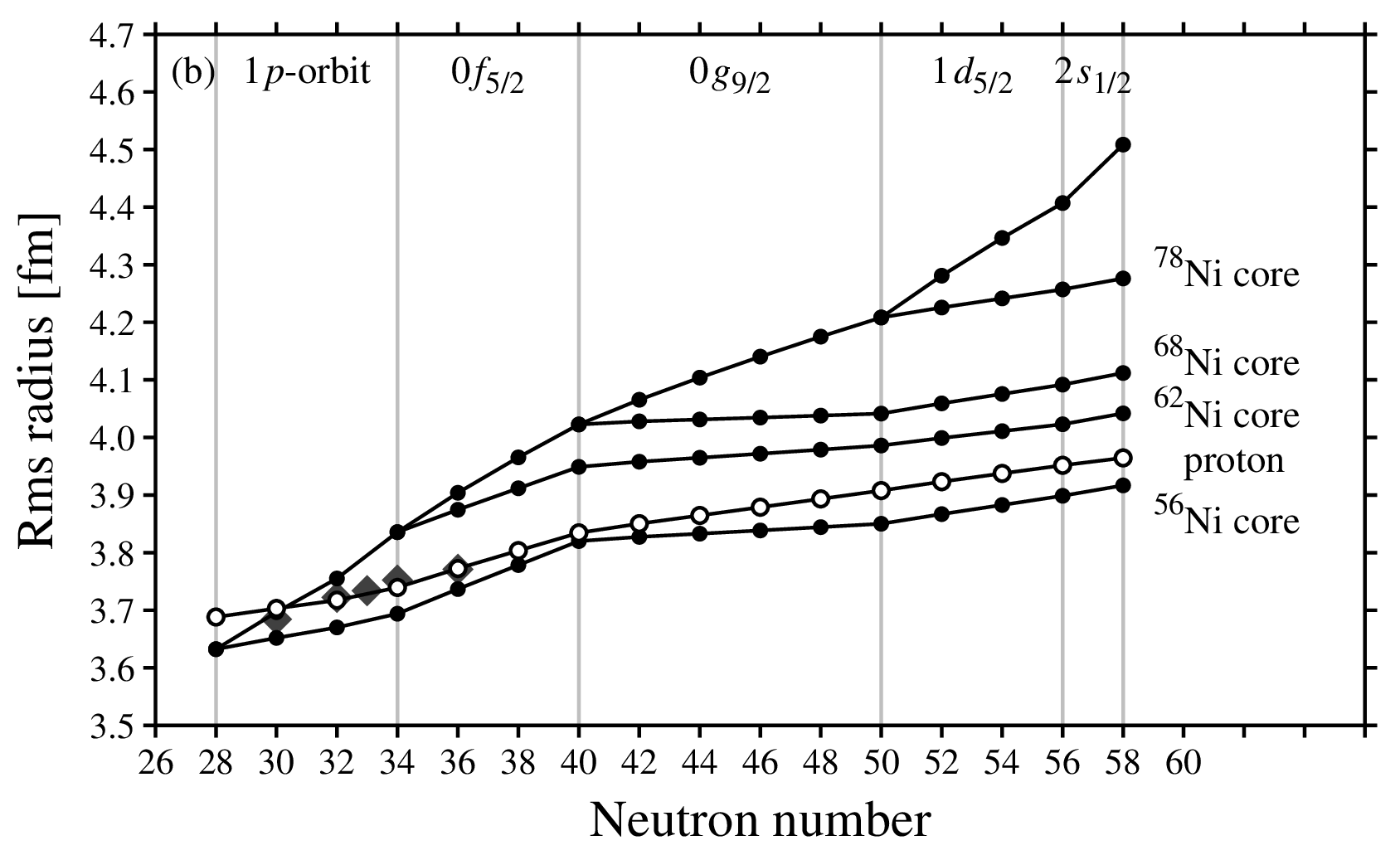, scale=0.3}
    \epsfig{file=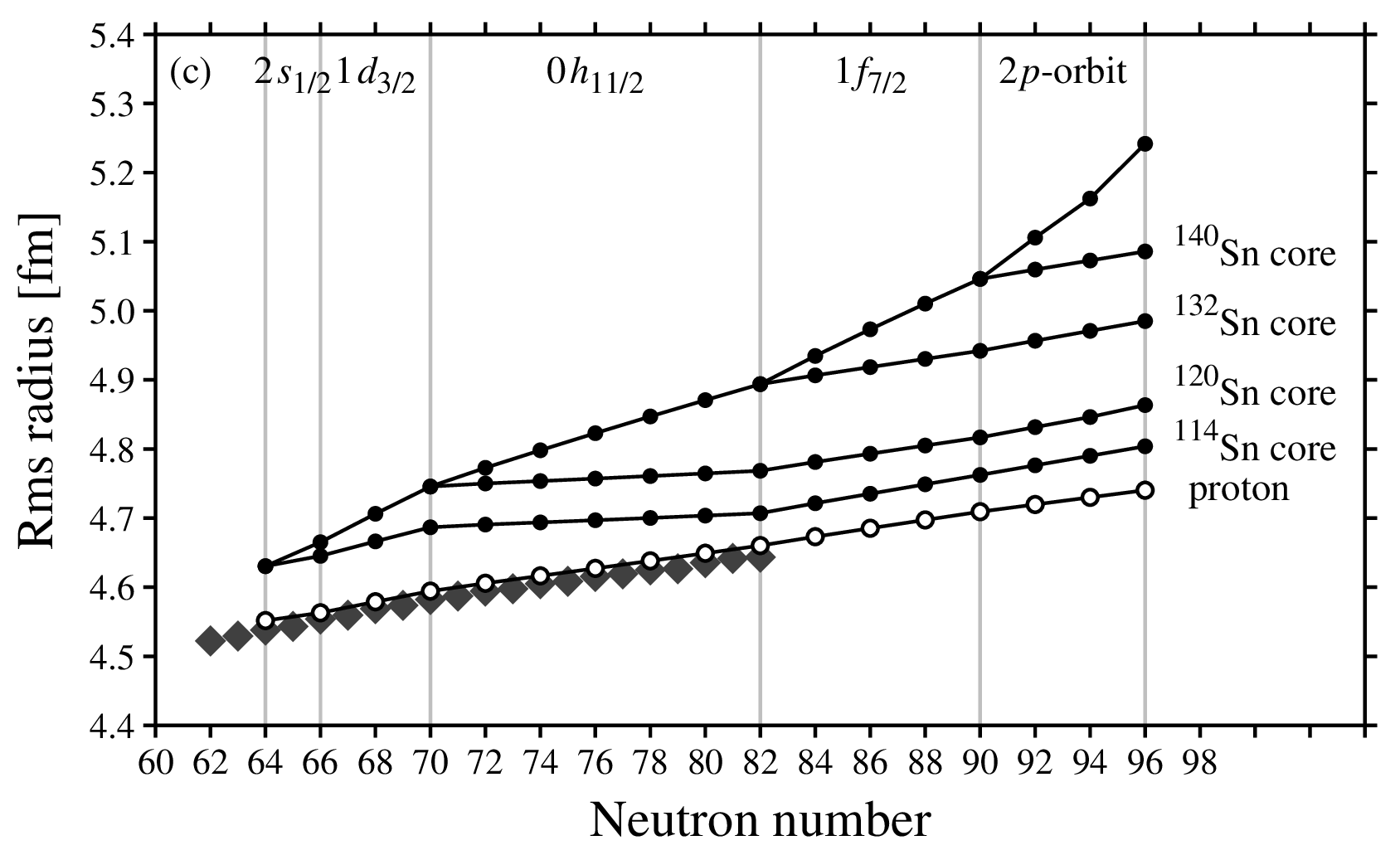, scale=0.3}
    \caption{Rms radii of (a) Ca, (b) Ni,
      and (c) Sn isotopes for neutron (closed circles)
      and proton (open circles).
      Rms neutron radii of various ``core'' nuclei are also plotted.
      The SkM$^\ast$ interaction is employed.
      See text for details.
      Filled diamonds in gray indicate
      experimental point-proton radii 
      extracted from Ref.~\cite{Angeli13}.
}
    \label{radius.fig}
  \end{center}
\end{figure}

Although there are small differences,
overall, the calculated densities reasonably
explain the behavior of the available experimental cross sections.
We discuss the size properties of the Ca, Ni, and Sn isotopes
using these ground-state wave functions obtained.
We calculate $^{40-70}$Ca ($N=20$--50), $^{56-86}$Ni ($N=28$--58),
and $^{114-146}$Sn ($N=64$--96)
to cover various single-neutron configurations across the major shells.
Here the ``core'' nucleus has to be defined in a
reasonable way so that we can discuss its size evolution.
For this purpose, we extract these single-particle orbits
that corresponds to the closed shell
or subshell neutron configurations from the full HF configurations.
The rms radii for neutrons and protons
are calculated for each ``core'' configuration
and their evolution with respect to the neutron number
is discussed.

Figure~\ref{radius.fig} draws the rms neutron and proton radii
of the Ca, Ni, and Sn isotopes obtained from the HF calculations 
and those extracted from their core configurations.
The results with the SkM$^\ast$ interaction are presented.
The rms point-proton radii extracted
from the charge radii~\cite{Angeli13} are also plotted
for comparison and are found to be in good agreement
with the theoretical calculations.
A kink behavior found at $N=28$ of the Ca isotopes
is qualitatively reproduced. Almost the same results are obtained
with the SLy4 and SkI3 interactions.
For more qualitative understanding
of the kink behavior of the nuclear radii,
we remark that the importance of the spin-orbit interaction
was discussed in Refs.~\cite{Nakada15a,Nakada15b,Nakada19}.
The kink behaviors of the proton radii at
$N=28$ in the Ca isotopes and at $N=126$ in the Pb isotopes
were well reproduced by introducing
density-dependent spin-orbit interaction.
The spin-orbit interaction could modify the single-particle
orbits and plays a role in the core swelling phenomena.

First, let us discuss the size evolution of the core nuclei
in the Ca isotopes drawn in Fig.~\ref{radius.fig} (a).
As the neutron number increases,
the total neutron radius gradually increases
showing some kink behavior at the neutron numbers
where the outermost single-neutron orbit is fully occupied.
In general, since the centrifugal barrier is smaller,
the lower angular momentum, the more diffused nuclear surface
obtains~\cite{Horiuchi17}, which leads to a larger rms radius.
On the other hand, the neutron number dependence of the core nuclei
behaves differently from that of the total neutron radii:
From $N=20$--28 and $N=40$--50,
the neutron radii of the $^{40}$Ca and $^{48}$Ca cores
are almost constant, whereas from $N=28$ to 40,
they increase drastically as it almost follows
the behavior of the proton radii.
This fact is also consistent with the observation found
in Ref.~\cite{Tanaka20}.
With large neutron excess at $N=40$--50,
the proton radius extends for gaining the symmetry energy
from the surrounding neutrons
despite the fact that the neutron radius keeps its constant behavior.
In fact, we see the linear enhancement
of the proton radius following the enhancement of the neutron radius
though the slope is not steeper than that of the neutron one.

In Fig.~\ref{radius.fig} (b),
similar behavior is also found in the Ni isotopes up to $N=50$
with further core swelling from $N=50$ to 58.
This isotopic chain extends to the Sn isotopes [Fig.~\ref{radius.fig} (c)],
showing the rapid increase of the neutron radius
from $N=64$--70 and $N > 82$,
whereas the proton radii gradually increase
without showing any kink behavior.

The enhancement of the neutron radius certainly
reflects the properties of the outermost single-particle orbits.
The constant behavior of the neutron core radius can only be found when
the nodeless $j$-upper orbits such as $0f_{7/2}$, $0g_{9/2}$,
and $0h_{11/2}$ orbits are occupied.
When the enhancement of the neutron core radius occurs,
the occupying orbits are always nodal wave functions
such as $1p, 1d, 2s, 1f_{7/2}, 2p$ orbits or $j$-lower orbits.
Since these single-particle densities
have strong overlap with the core density,
the nuclear density in the internal region becomes high
when nodal and $j$-lower orbits are added to the core nucleus,
while the nodeless $j$-upper orbits only
contribute to the densities at around the nuclear surface.
If the internal density of the core is already saturated,
the system looses the energy, and thus,
to accommodate the additional neutrons,
strong rearrangement of the mean field occurs 
to decrease the density in the internal region,
which leads to the core swelling.
We also made the same analysis for $^{16-24}$O.
As expected from the above conjecture, the swelling of the $^{16}$O core
occurs from $N=14$--16 where the $1s_{1/2}$ orbits contribute,
which is consistent with the consequence given in Ref.~\cite{Masui12}.

\begin{figure*}[ht]
  \begin{center}
    \epsfig{file=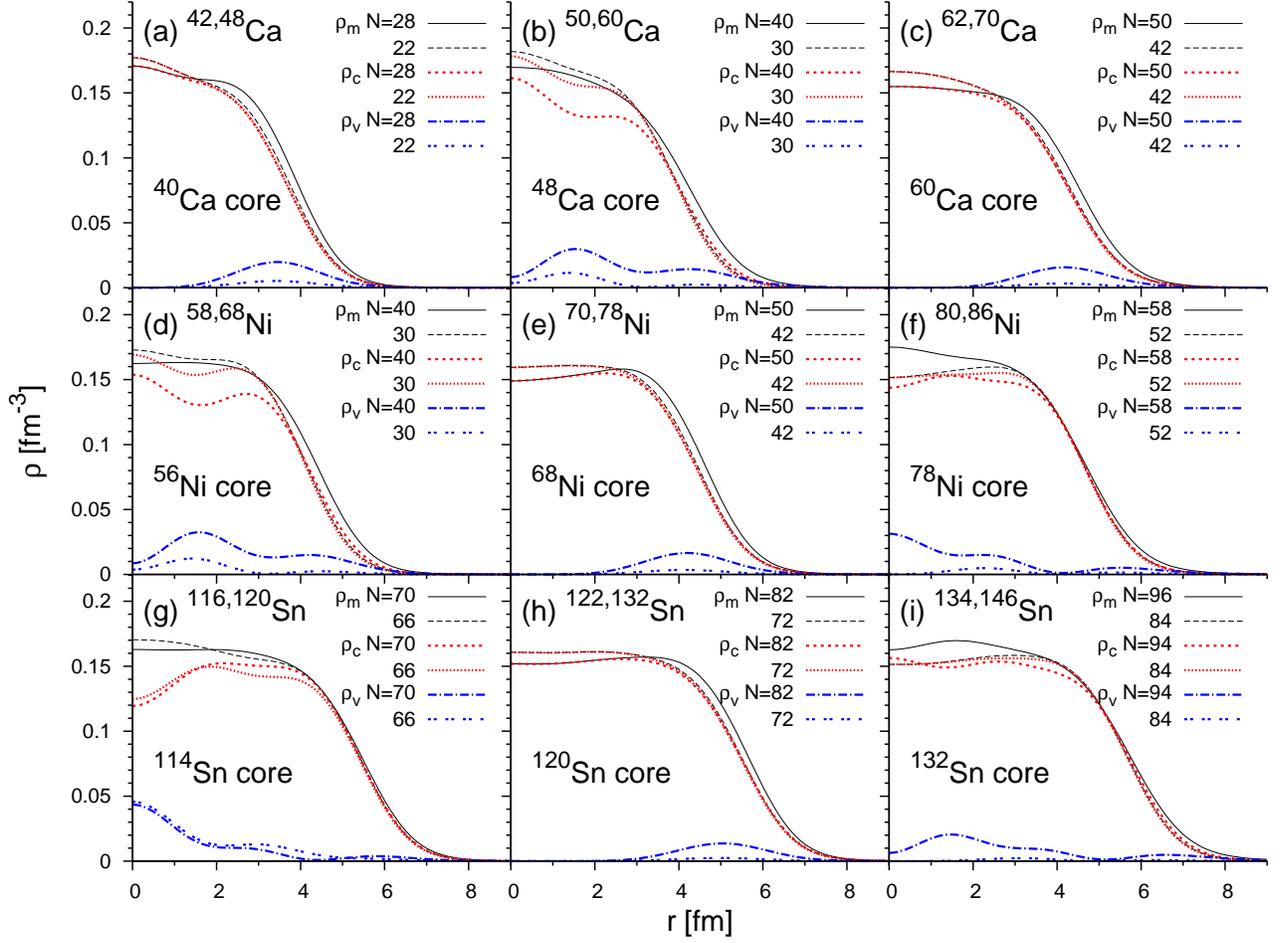, scale=1.1}    
    \caption{Total matter ($\rho_m$), core ($\rho_c$),
      and valence neutron ($\rho_v$) densities
      of $^{42,48,50,60,62,70}$Ca, $^{58,68,70,78,80,86}$Ni,
      and $^{116,120,122,132,134,146}$Sn. The SkM$^\ast$ interaction is employed.
      See text for details.}
    \label{density.fig}
  \end{center}
\end{figure*}

\begin{figure}[ht]
  \begin{center}
    \epsfig{file=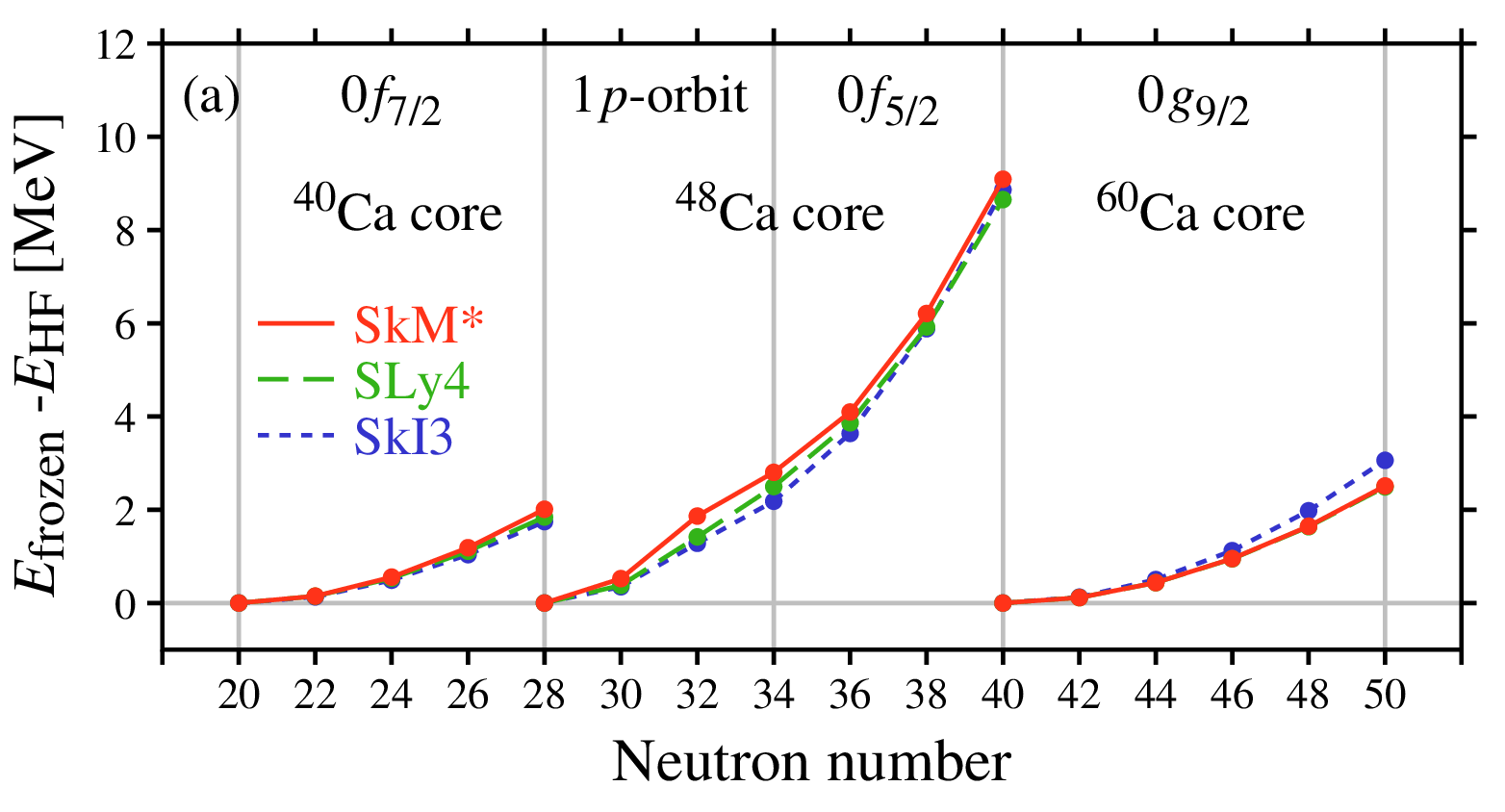, scale=0.35}
    \epsfig{file=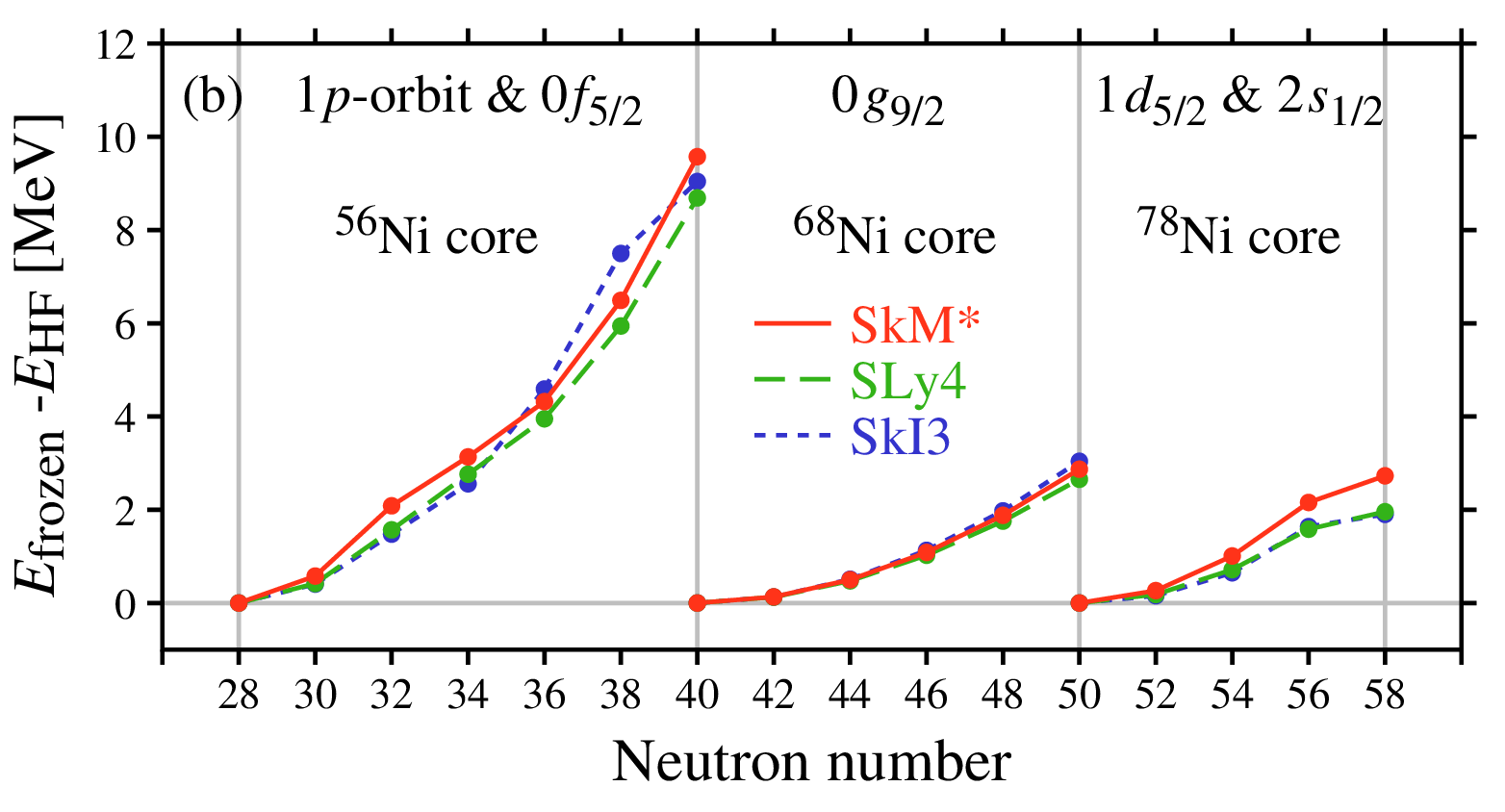, scale=0.35}
    \epsfig{file=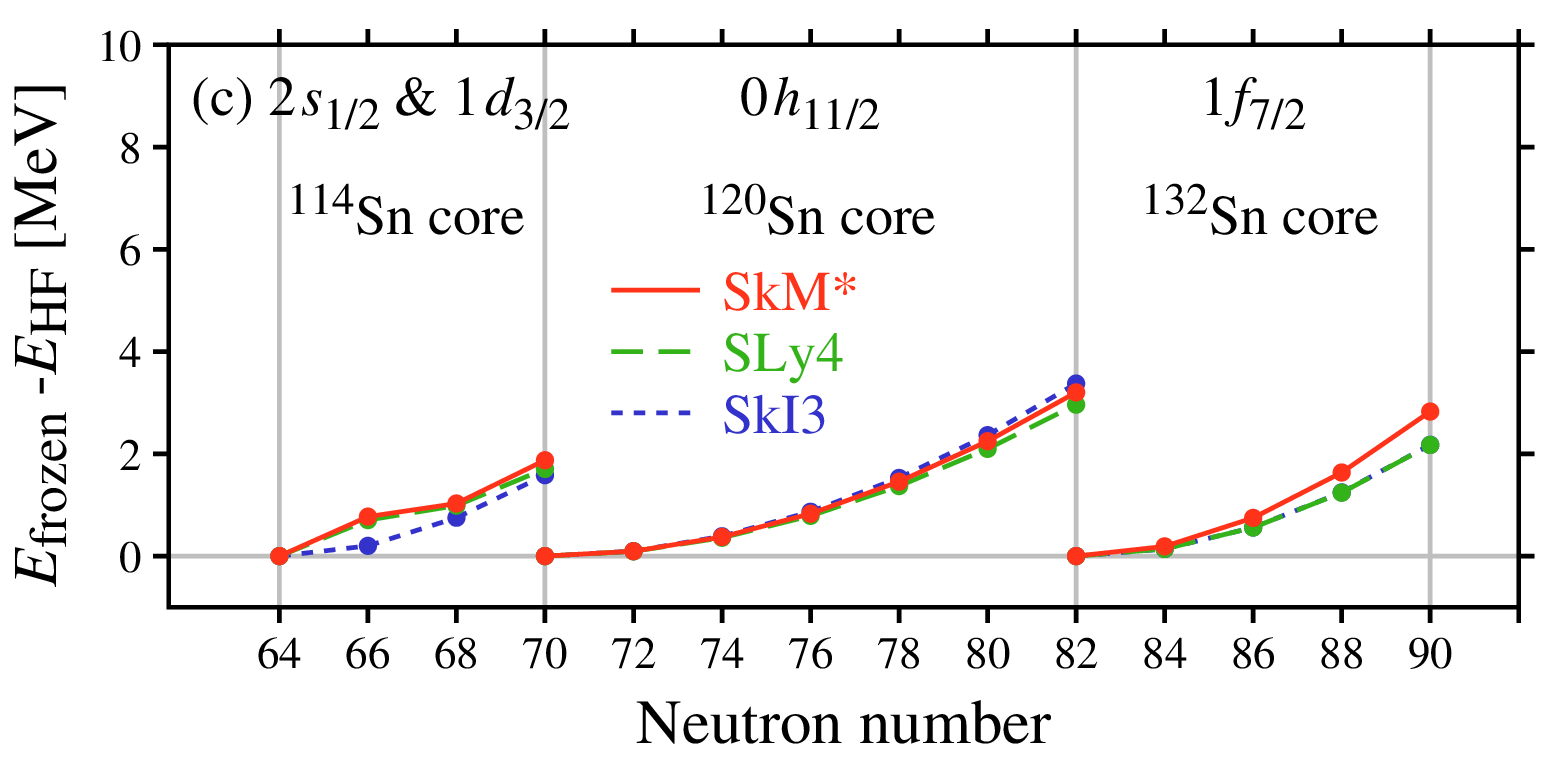, scale=0.35}
    \caption{Energy difference between the HF and the frozen-core HF
      calculations of (a) Ca, (b) Ni, and (c)
      Sn isotopes with the SkM$^\ast$, SLy4, and
      SkI3 interactions. See text for details. }
    \label{energy.fig}
  \end{center}
\end{figure}

To show the above finding more explicitly, in Fig.~\ref{density.fig},
we display the total matter (solid and long-dashed lines in black),
core (dashed and dotted lines in red),
and valence neutron densities (chain and double-dotted-spaced lines in blue)
of the selected Ca, Ni, and Sn isotopes located in different
shell- or subshell regions.
In the regions where the core swelling do not occur,
which correspond to
the panels (a), (c), (e), and (h) of Fig.~\ref{density.fig},
the valence neutrons (chain and double-dotted-spaced lines)
occupy the $j$-upper orbits and contribute mainly to the enhancement
of the densities of the nuclear surface at around the nuclear half density.
In case of $^{50}$Ca and $^{58}$Ni ($N=30$), the panels (b) and (d),
respectively, we see that the central densities of the core
(dotted line) are already saturated.
The valence neutrons (double-dotted-spaced lines)
occupy the nodal or $j$-lower orbits and
contribute to the densities of all regions.
By adding neutrons up to $N=40$,
the central densities of the $^{48}$Ca or $^{56}$Ni core (dashed lines)
have to be reduced to keep the saturation density $\sim 0.16$--0.17 fm$^{-3}$
in the total matter densities (solid lines).
Thus the core densities are enhanced in the surface region.
In the panels (f), (g), and (i),
the central densities of the core (dashed and dotted lines)
becomes $\sim 0.15$ fm$^{-3}$,
which can accommodate further neutrons as they are indicated
in the valence neutron densities (chain and double-dotted-spaced lines).
In these isotopes, the matter radii
are enhanced due to the contributions from
both the core and valence neutron densities at the surface region.
The internal nucleons are pulled by
the excess neutrons to gain the symmetry energy
as presented in the total matter densities
(solid and long-dashed lines) in the panels (e) and (h).

To strengthen the above conjecture,
we perform another HF calculation for the valence neutrons
by assuming an inert or frozen core in the system. 
Actually, we evaluate the energy difference between
the HF calculation $E_\mathrm{HF}$ and the HF calculation
with frozen core, $E_\mathrm{frozen}$.
In the frozen-core HF calculations,
the wave function of the ``core'' nucleus
is replaced by the ground-state wave
function of the corresponding nucleus.
For instance, the $E_\mathrm{frozen}$ of $^{52}$Ca,
which is assumed to have $^{48}$Ca core and four valence neutrons,
is calculated by optimization of the four valence neutron wave functions.
The wave functions of the $^{48}$Ca core is always represented by 
the ground state of $^{48}$Ca obtained by the full HF calculation. 
No rearrangement among core and valence neutron
configurations occurs in the frozen-core HF calculations.
Therefore, the energy difference $E_\mathrm{frozen} - E_\mathrm{HF}$ can be
a measure of the mean-field rearrangement by adding the valence neutrons.

Figure~\ref{energy.fig} displays
$E_\mathrm{frozen} - E_\mathrm{HF}$ of the Ca, Ni, and Sn isotopes.
No qualitative difference is found in the calculations with
the SkM$^\ast$, SLy4, and SkI3 interactions.
As expected, the energy loss appears to be large
when the nodal or $j$-lower orbits are occupied,
especially, in $N=28$--40 for the Ca and Ni isotopes.
On the other hand,
the energy difference is small when the nodeless $j$-upper orbits,
the $0f_{7/2}$, $0g_{9/2}$, and $0h_{11/2}$ orbits, are filling,
in which no swelling the neutron core occurs as displayed
in Fig.~\ref{radius.fig}. This means that in those $N$ regions
the core and valence neutrons densities are
well separated in terms of the nuclear saturation.

We find a moderate energy loss in 
$N=50$--58 for the Ni isotopes and $N=64$--70, 82--96 for the Sn isotopes
despite the fact that a weak increase of the neutron core radius occurs
as presented in the panels (b) and (c) of Fig.~\ref{radius.fig}.
This is because the internal densities of
those core nuclei, $^{60}$Ca and $^{114,132}$Sn,
are not well saturated as presented in 
in the panels (f), (g), and (i) of Fig.~\ref{density.fig} (dashed lines)
for $N=58$, $N=70$, and $N=84$, respectively.
Since these internal densities are low enough to accommodate
the additional neutrons that comes from the internal amplitudes
of the nodal orbits, the rearrangement of
the mean field is weaker, and thus the energy loss is suppressed.

{\it Conclusions--}
The sudden increase of the nuclear radii is the indication
of the saturation of the internal density balanced with
the nuclear symmetry energy.
It is found that the evolution
of the neutron radius is strongly depend
on the occupying valence neutrons.
By adding the nodeless $j$-upper orbits, 
it shows the constant behavior of the neutron radii of the core nucleus
because these orbits only contribute to the nuclear surface
where the nuclear densities are unsaturated.
A sudden increase of the neutron core radius occurs
when the nodal or $j$-lower orbits are occupied.
Strong rearrangement of the mean field happens to reduce
the central density to accommodate these orbits
that contribute to the densities of all regions.
This behavior is significant in light-medium mass nuclei,
such as Ca and Ni isotopes as the valence nucleons
strongly influence the core configurations.
For heavier nuclei, Sn, the neutron and proton radii are determined in
a more democratic way.
Since more neutrons can be accommodated
in the internal region,  the moderate core swelling is expected
for such nuclei with large neutron excess,
where the internal densities are not well saturated.

This knowledge is also useful when one formulates
a core plus few-nucleon model.
When the valence nucleons occupy the nodal or $j$-lower orbits,
some energy loss is expected if the core nucleus is inert or
the core swelling is not introduced. However,
this effect may be small
in the case of the dripline nuclei because
the core nucleus is expected to be already swelled and
its internal density is unsaturated.

Systematic studies of the neutron and proton radii
and densities are desired to reveal
these nuclear size properties of finite nuclei.
For this purpose, a systematic measurement of
the total reaction cross sections
on a proton target~\cite{Horiuchi14,Horiuchi16}
may be the most promising way to investigate
the core swelling phenomena for the short-lived
neutron-rich nuclei.

In this paper, we only investigate the spherical nuclei for
simplicity. A further study considering the nuclear deformation
is interesting and will be reported elsewhere.

{\it Acknowledgments--} This work was in part supported
by JSPS KAKENHI Grants Nos.\ 18K03635, 18H04569, and 19H05140.
We acknowledge the collaborative research program 2019, 
information initiative center, Hokkaido University.

\end{document}